\documentclass[sigconf, nonacm]{acmart}

\usepackage{booktabs} 
\usepackage{amsmath} 
\usepackage{graphicx} 
\usepackage{caption} 
\usepackage{subcaption} 
\usepackage{float}
\usepackage{placeins} 
\usepackage{afterpage}

\acmConference[ACM NanoCom 2024]{Proceedings of the 11th ACM International Conference on Nanoscale Computing and Communication}{October 28-30, 2024}{Milan, Italy}

\begin{document}
\title{Experimental Characterization of Hydrodynamic Gating-Based Molecular Communication Transmitter}

\author{Eren Akyol$^{\ast}$ \hspace{2mm}Ahmet Baha Ozturk$^{\ast \dag}$ \hspace{2mm}Iman Mokari Bolhassan$^{\ast}$ \hspace{2mm}Murat Kuscu$^{\ast \ddag}$}
\affiliation{%
  \institution{$^{\ast}$ Nano/Bio/Physical Information and Communications Laboratory (CALICO Lab), Koç University, Türkiye \\
  $^{\dag}$ Istanbul Technical University, Türkiye \\
  $^{\ddag}$ Center for neXt-generation Communications (CXC), Koç University, Türkiye \\ $^{\ddag}$ Nanofabrication and Nanocharacterization Center for Scientific and Technological Advanced Research (n\textsuperscript{2}STAR), Koç University, Türkiye \\ $^{\ddag}$ Koç University Research Center for Translational Medicine (KUTTAM), Türkiye}
  \city{}
  \country{}
}
\email{eakyol22@ku.edu.tr, ozturkahm21@itu.edu.tr, ibolhassan22@ku.edu.tr,mkuscu@ku.edu.tr}

\renewcommand{\shortauthors}{Akyol et al.}

\begin{abstract}
Molecular communication (MC) is a bio-inspired method of transmitting information using biochemical signals, promising for novel medical, agricultural, and environmental applications at the intersection of bio-, nano-, and communication technologies. Developing reliable MC systems for high-rate information transfer remains challenging due to the complex and dynamic nature of application environments and the physical and resource limitations of micro/nanoscale transmitters and receivers. Microfluidics can help overcome many such practical challenges by enabling testbeds that can replicate the application media with precise control over flow conditions. However, existing microfluidic MC testbeds face significant limitations in chemical signal generation with programmable signal waveforms, e.g., in terms of pulse width. To tackle this, we previously proposed a practical microfluidic MC transmitter architecture based on the hydrodynamic gating technique, a prevalent chemical waveform generation method. This paper reports the experimental validation and characterization of this method, examining its precision in terms of spatiotemporal control on the generated molecular concentration pulses. We detail the fabrication of the transmitter, its working mechanism and discuss its potential limitations based on empirical data. We show that the microfluidic transmitter is capable of providing precise, programmable, and reproducible molecular concentration pulses, which would facilitate the experimental research in MC.
\end{abstract}

\keywords{Microfluidics, Molecular Communication, Pulse Shaping, Hydrodynamic Gating} 

\maketitle

\begin{figure*}
    \centering
    \includegraphics[width=\linewidth]{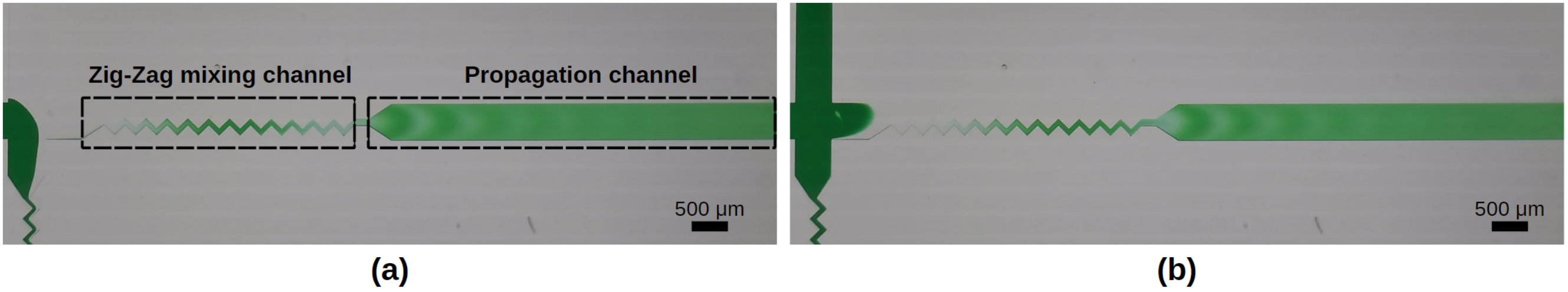}
\caption{An illustration of different states of the hydrodynamic gating process: (a) Gating state (barricading the entrance of information molecules), (b) Injection state (allowing the entrance of information molecules). Note: The zig-zag mixing channel in this chip, which shows a single transition from left to right, is extended in subsequent designs to increase the hydraulic resistance by having multiple transitions.
}

    \label{fig:onoff}
\end{figure*}

\section{Introduction}

Interfacing human-made micro/nanoscale devices with natural biological systems, through molecular information exchange, conceptualized as Internet of Bio-Nano Things (IoBNT), is promising for groundbreaking applications, such as continuous health monitoring and targeted drug delivery \cite{akyildiz2015internet, kuscu2021internet, akan2023internet}. IoBNT necessitates unconventional communication techniques to interconnect such \emph{bio-nano things} of diverse nature \cite{kuscu2021internet}. A promising approach is bio-inspired molecular communications (MC), where information is encoded and transmitted through the exchange of molecules. MC offers inherent biocompatibility, energy efficiency, and reliability in physiological and confined environments, where conventional communication techniques, e.g., electromagnetic communications, are inefficient, unsafe, or not feasible.  \cite{akan2016fundamentals, nakano2012molecular, farsad2016comprehensive}.

Following a substantial body of theoretical research \cite{jamali2019channel, kuscu2019transmitter, akyildiz2019information}, many experimental studies on MC have been conducted using testbeds of various scales, media, and complexities. Readers can refer to \cite{10105650, 10102773} for a comprehensive review of recent experimental works in MC. Common to many of the microscale MC testbeds is the use of microfluidic technologies, which enable precise control over channel geometries and flow conditions, essential for replicating realistic application environments \cite{hamidovic2024microfluidic, walter2023real}. While the propagation channels in these microfluidic testbeds have significant commonalities, the transmitter and receiver architectures vary largely, determining the range of MC scenarios that can be simulated with these testbeds. Although MC receiver designs compatible with microfluidic channel geometries, particularly lateral flow sensor-based architectures, have been widely studied \cite{abdali2024frequency, kuscu2021fabrication, martins2022microfluidic, kuscu2018modeling}, the design of transmitters and pulse-shaping techniques for these microfluidic testbeds has received relatively less attention \cite{zadeh2023microfluidic}.

A limited number of microfluidic MC transmitters have been proposed in the literature. For instance, Bi et al. utilized flow- and geometry-controlled chemical reactions in microfluidic channels to generate predefined molecular concentration pulses in response to rectangular triggering signals \cite{bi2020chemical, bi2022microfluidic}. Guo et al. developed signal generators that produce multiple concentration pulses with four different amplitudes, inherently functioning as microfluidic MC transmitters \cite{guo2022time, guo2020multichannel}. Their work underscores the importance of considering signal amplitude as a key property in the design of microfluidic transmitter devices. However, despite advancements, these devices often face significant challenges, such as the lack of complete control over signal properties, particularly pulse width, hindering their adaptability and usability. Additionally, existing devices can be complex and costly, limiting their widespread adoption.

To address previous challenges, we proposed a practical microfluidic MC transmitter using a hydrodynamic gating technique in \cite{10418566} with high spatiotemporal resolution. We developed an approximate analytical model to capture the fundamental characteristics of generated molecular pulses, such as pulse width, amplitude, and delay, as functions of main system parameters such as flow velocity and gating duration. We validated the accuracy of this model by comparing it with finite element simulations using COMSOL Multiphysics under various system settings.

In this paper, we experimentally validate and characterize the performance of the hydrodynamic gating-based MC transmitter. The previous MC transmitter design was modified to enhance control over generated signals, particularly pulse width. Our focus in this new design is also on enhancing reproducibility in the cross-sectional distribution of information molecules and improving the predictability of the transmitted molecular pulses. To this end, a zig-zag mixing structure is integrated into the transmitter architecture.

The paper is organized as follows: Section \ref{sec:transmitter design} details the architecture of the transmitter and its working mechanism. In Section \ref{sec:fabrication}, we describe the fabrication processes of our microfluidic MC transmitter chip. Section \ref{sec:experimental} describes the experimental setup, and discusses the data processing techniques utilized. Finally, Section \ref{sec:Conclusions} concludes the paper with a brief discussion of future work.

\section{Microfludic MC Transmitter Design}
\label{sec:transmitter design}
\begin{figure}[h]
    \centering
    \includegraphics[width=\linewidth]{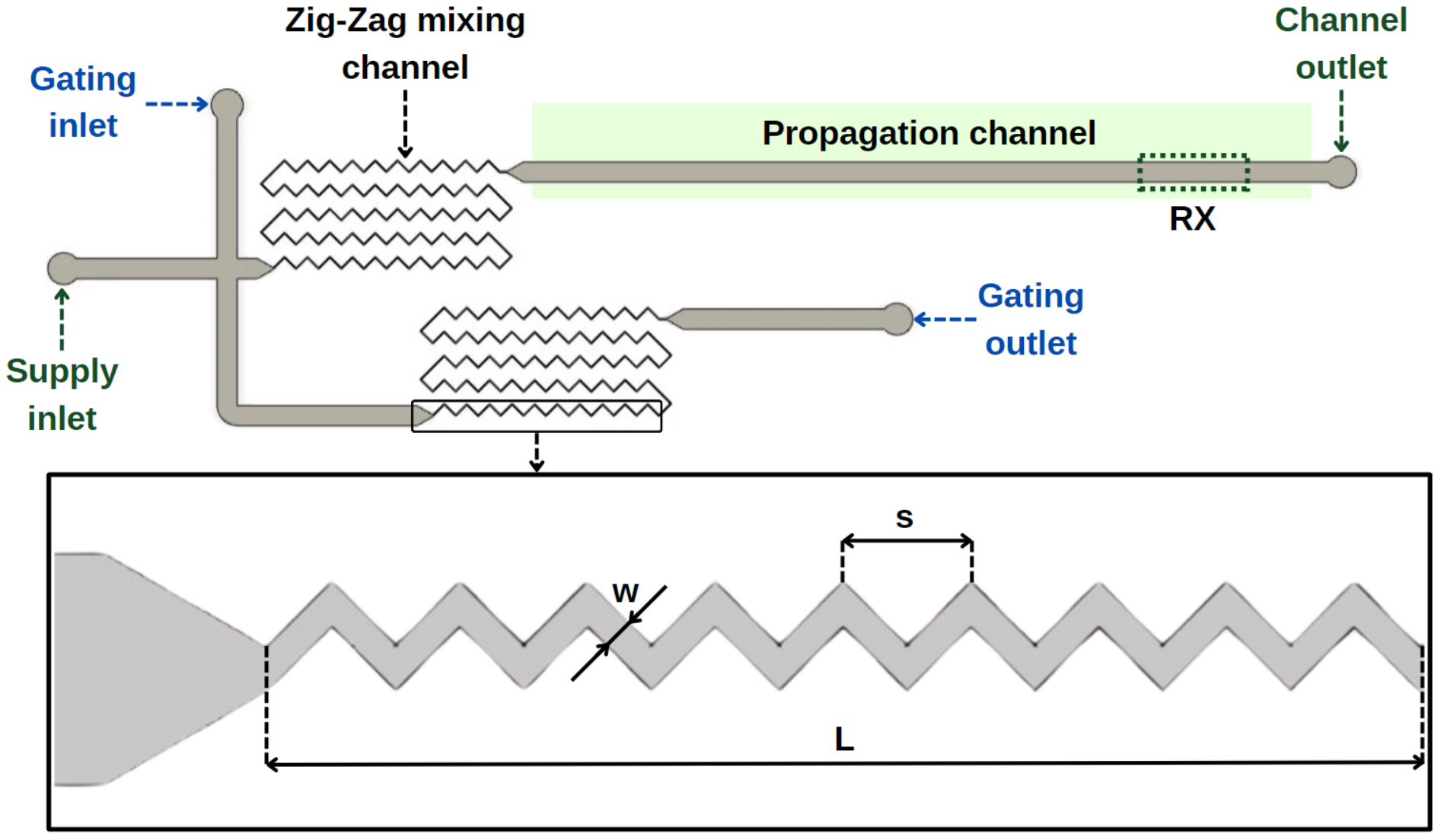}
    \caption{Schematic representation of the Hydrodynamic Gating-Based MC Transmitter and parameters of the zig-zag mixing sturcture. $L$ is the total linear length, $s$ is the length of each periodic zig-zag step, and $w$ is the channel width.}
    \label{fig:design}
\end{figure}

To implement the hydrodynamic gating-based microfluidic MC transmitter, we built upon the cross-shaped transmitter design from our previous work \cite{10418566}, introducing several modifications as shown in Fig. \ref{fig:onoff} and Fig. \ref{fig:design}. The modified design includes a mixing structure between the center of the cross-shape and the outputs to ensure a uniform mixture, inspired by Shao et al. \cite{shao2023cross}, who used electroosmotic drive for dynamic control in a cross-channel microfluidic device. This structure allows dynamic control of input ratios in terms of volumetric flux, ensuring a homogeneous cross-sectional distribution of the mixture fluid. Identical mixing structures in both output channels maintain symmetry, preventing flow inconsistencies and directing flux evenly, avoiding resistance imbalances \cite{oh2012design}. By ensuring uniformity, we achieve consistent fluid behavior and accurate signal generation across both channels. For a simpler mechanism, we utilized the pressures of the two inputs for dynamic control.

Our choice of the appropriate type of mixing structure was based on what seemed to be relatively easier to manufacture and sufficient in terms of mixing efficiency. Passive mixers were preferred over active mixers as they did not require additional physical mechanisms, which was in agreement with our goal of achieving a simpler mechanism. Among passive mixers, the zig-zag structured channels seemed to fit our criteria as they both had relatively high mixing efficiency and were simpler in terms of manufacturing, as some of the other structures, such as 3-D serpentine structures, create greater difficulty in manufacturing compared to 2-D structures \cite{lee2011microfluidic}. Hence, the zig-zag structure was chosen for its ease of design and fabrication, and effective mixing performance, and was deemed the appropriate mixing structure. Apart from the mixing structure type, the effect of the channel width on mixing efficiency was also researched, and the common conclusion that narrower channels led to proportionally more efficient mixing was utilized by designing the zig-zag channel widths to be around $0.1$ times the original channel length \cite{knight1998hydrodynamic}.

After choosing the zig-zag structure, parametric studies were researched to determine the appropriate ratios of the zig-zag in comparison to the original channel geometries \cite{mengeaud2002mixing}. These parameters were mainly the channel width ($w$), the total linear length of the zig-zag microchannel ($L$), and the linear length of each individual segment or "step" in the zig-zag pattern ($s$). Here, $s$ refers to the horizontal length of each segment between the points where the channel changes direction. It was concluded that, regardless of the mixing structure, the mixing efficiency was proportional to the total linear length of the microchannel and inversely proportional to its width. Hence, the research on the parameters focused on the zig-zag structure's unique parameters, such as the ratio of $s/w$. It was then further concluded that the only parameter worth considering was the ratio of $s/w$. This was because the geometric parameters of $w$ and $L$ were considered to be constants. Hence, the ratio of $w/L$ is constant. This leaves us with the $s$ parameter to be studied, which can be done by considering the ratio of either $s/w$ or $s/L$, where either would be sufficient in determining the optimal $s$ value due to the constant values of $w$ and $L$. In reference to the parametric study \cite{mengeaud2002mixing}, the result that the ratio value $s/w  = 4$ provides maximum mixing efficiency was utilized within our design.

After determining the design, we established the operational aspects of our microfluidic transmitter. To generate sequential pulses, we utilized a cross-shaped architecture and the hydrodynamic gating method. Our transmitter operates in two main states: gating ON mode and gating OFF mode. In the gating ON mode, depicted in Fig. \ref{fig:onoff} (a), the propagation of pulses to the mixing compartment is obstructed by the dominance of the gating inlet. The time interval for the gating ON mode is maintained to ensure the desired inter-pulse duration. The entrance of the information carrier molecules, which in our case are simply green food dye, into the mixing structure occurs in the gating OFF mode, illustrated in Fig. \ref{fig:onoff} (b). To achieve precise control over inter-pulse duration and pulse width of the signals, we applied some pressure to the gating inlet, rather than the common approach of setting the gating pressure to zero. This pressure ensures that the gating molecules admit the entrance of the information carriers.

\section{Fabrication}
\label{sec:fabrication}
\begin{figure}
    \centering
    \includegraphics[width=\linewidth]{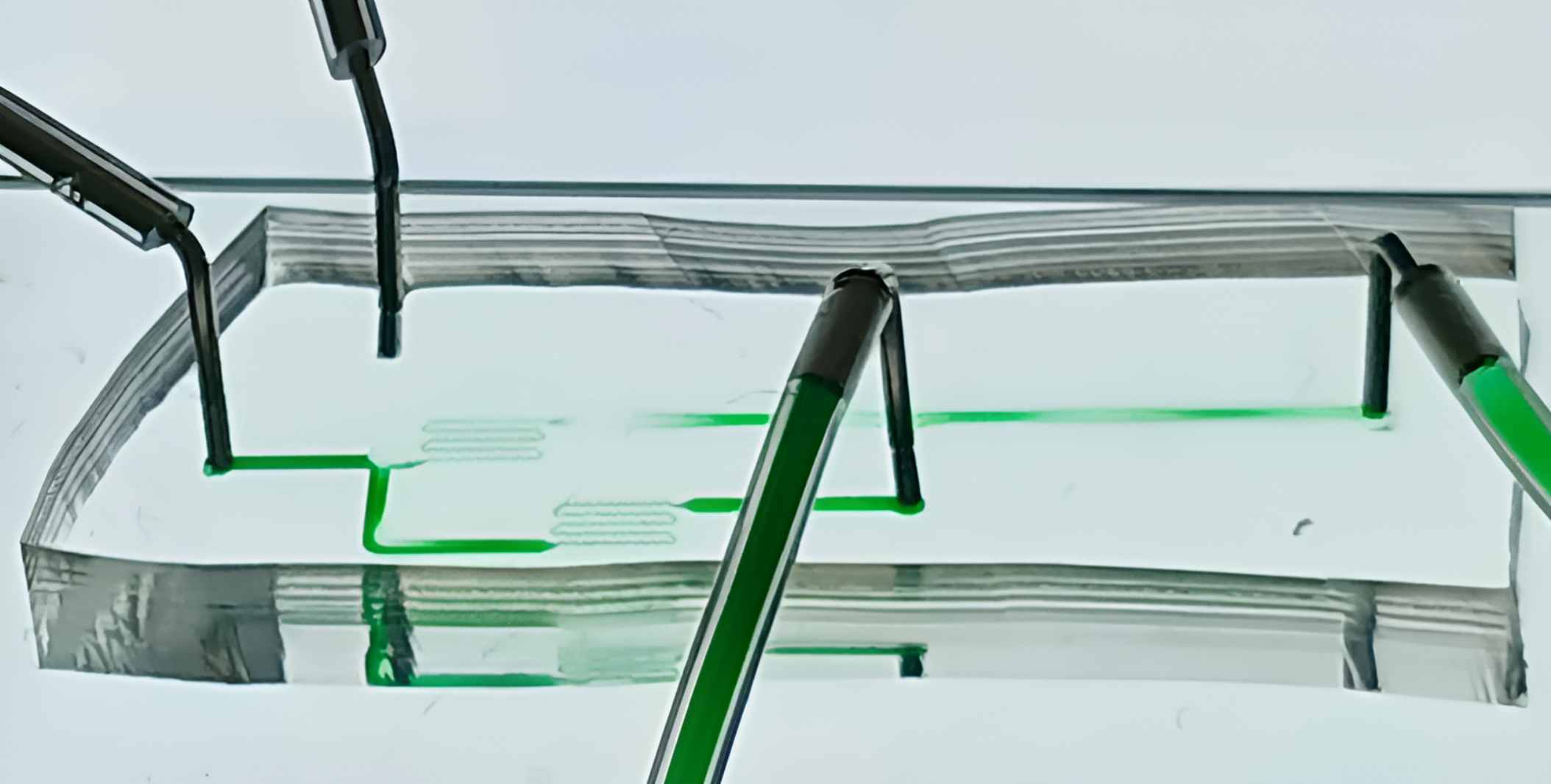} 
    \caption{An illustration of the fabricated PDMS chip for the MC transmitter, integrated to the experimental setup.}
    \label{fig:fabrication}
\end{figure}

The fabrication process of the MC transmitter chip, shown in Fig. \ref{fig:fabrication}, involves three main steps: mold fabrication, PDMS preparation and casting, and bonding. The mold is created by spin-coating a 4-inch silicon oxide wafer with SU-8 3050 photoresist at 2000 rpm for 90 seconds. After a soft-bake at 60°C for 2 minutes and 90°C for 7 minutes, the wafer is exposed to UV light using a maskless lithography system (Heidelberg MLA 100) to define the pattern. A post-bake follows at 60°C for 3 minutes and 90°C for 8 minutes. The wafer is then developed in SU-8 developer solution for 15 minutes to dissolve the unexposed photoresist. For PDMS preparation, Sylgard 184 Silicone Elastomer is mixed with a curing agent in a 10:1 mass ratio and degassed to remove air bubbles. The mixture is poured onto the mold and cured at 70°C for one hour in a vacuum oven. The cured PDMS is peeled off, cut into individual chips, and punched with inlet and outlet holes. In the final step, PDMS chips and glass slides are cleaned with isopropanol (IPA), treated with oxygen plasma, and then bonded together.

\FloatBarrier

\section{Experimental Work}
\label{sec:experimental}
\subsection{Experimental Setup}
To supply precise, reproducible, and independently generated pressure inputs at the two inlets of the transmitter, we used a commercial pressure controller system (OB1 MK3 - Microfluidic flow control system, Elveflow). The pressure inputs were connected to two separate closed reservoirs, where one contained green food dye and the other contained tap water. The choice of these fluids was due to the high imaging contrast they provide for the generated pulses during the succeeding video processing steps in MATLAB. Microscopic imaging and video recording were performed at 60 FPS using a 3D microscope (Microqubic MRCL700) in 2D inverted configuration. To prevent measurement artifacts caused by varying reservoir heights affecting inlet pressure, we fixed the positions and heights of all components, including the chip and reservoirs. In the next step, the reservoirs containing the food dye and water are connected to the chip inlets via tygon tubes and metallic tips. Another tygon tube is connected to the outlet for transfering the fluid to a waste container. After that, we provide the corresponding pressure values for gating ON mode to ensure that our transmitter reaches equilibrium and is ready to use. The pressure controller is then programmed to apply periodic pressure pulses to the water reservoir with an offset pressure value, such that a series of gating ON and OFF modes, as briefly explained in Section \ref{sec:transmitter design}, is realized, generating successive concentration pulses with the desired inter-pulse durations and input pulse widths. 

\begin{figure*}[h]
    \centering
    \includegraphics[width=\linewidth]{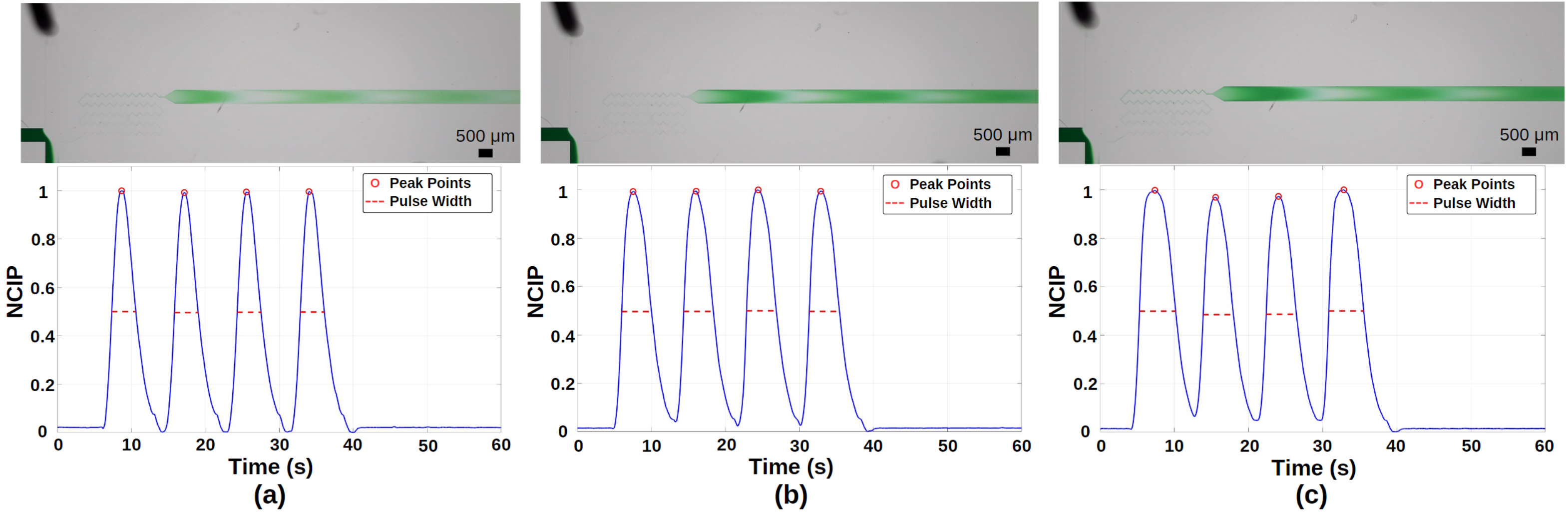}
    \caption{An illustration of consecutive pulses with different gating times ($T\textsubscript{g}$): (a) t = 90 ms, (b) t = 120 ms and (c) t = 150 ms. The dashed lines indicate the pulse width values ($w_{p}$), determined using the FWHM method, while the peak points mark the maximum concentration points. NCIP stands for normalized concentration intensity profile. Processed with MATLAB.}
    \label{fig:PW}
\end{figure*}

\subsection{Results and Analysis}
In MC systems, there are several parameters that directly have an influence on the data transmission rate and overall performance of the system. In the explained experimental setup, we focused on two crucial characteristics of signals: pulse width, $w\textsubscript{p}$ and measured inter-pulse duration, $T_\mathrm{pm}$. 

Although minimizing $T_\mathrm{pm}$ and $w_\mathrm{p}$ increases the data transmission rate in theory, there exist particular challenges in MC due mainly to inter-symbol interference (ISI), which complicates the detection of transmitted concentration pulses. In the first part of the experiments, we investigated the precision of the transmitter in generating pulses with different $w_\mathrm{p}$, providing that input inter-pulse duration, $T_\mathrm{pi}$, are set to $8$ seconds. In the second part, we explored the limits of the device by observing the effects of ISI for varying $T\textsubscript{pi}$, provided that the gating time, $T_\mathrm{g}$ is $100$ milliseconds. The pressures of the two inlets for the two states of the transmitter, gating ON and OFF modes, are maintained the same throughout the experiments to minimize the potential effects of velocity variations.

\begin{table}[H]
    \centering
    \caption {Relevant Parameters of the MC Transmitter}
    \label{table:parameters}
    \scalebox{0.9}{
    \begin{tabular}{|c|p{6cm}|}
        \hline
        \textbf{Parameter} & \multicolumn{1}{c|}{\textbf{Description}} \\
        \hline
        \texttt{$w_{p}$} & \multicolumn{1}{c|}{Pulse width} \\
        \hline
        \texttt{$T_{pm}$} & \multicolumn{1}{c|}{Measured inter-pulse duration} \\
        \hline
        \texttt{$T_{pi}$} & \multicolumn{1}{c|}{Input inter-pulse duration} \\
        \hline
        \texttt{$T_{g}$} & \multicolumn{1}{c|}{Gating time} \\
        \hline
    \end{tabular}
    }
\end{table}

The experimental investigation aimed to assess the performance of our transmitter chip across a range of $T_\mathrm{g}$ spanning from 80 ms to $200$ ms with 10 increments. To accurately measure $w_\mathrm{p}$, we have used the Full Width at Half Maximum (FWHM) metric. Pulse width $w_\mathrm{p}$ is a crucial parameter because it provides insights into the temporal characteristics of signal transmission, affecting the duration of information exchange and determining the spatial extent of signal propagation. In Fig. \ref{fig:PW}, the observations for three different $T_\mathrm{g}$ values —$90$, $120$, $150$ ms— at the starting point of the propagation channel are provided to illustrate our transmitter’s ability to generate signals with varying characteristics. To determine $w_\mathrm{p}$ corresponding to each $T_\mathrm{g}$ value, we calculate $w_\mathrm{p}$ for each consecutive pulse and average the $w_\mathrm{p}$ values of the transmitted pulses. The measurement results for $w_\mathrm{p}$ across three different $T_\mathrm{g}$ values reveal values of $3.28$, $4.07$, and $4.7$ seconds. These results establish a linear increase in $w_\mathrm{p}$ as the gating time $T_\mathrm{g}$ increases, detailed analysis are provided in Fig. \ref{fig:tables} (b). As a result, the graphs in Fig. \ref{fig:PW} demonstrate that our transmitter produces stable and repeatable results across different $T_\mathrm{g}$ values underlining its potential to improve the reliability and reproducibility of MC systems.

An important limitation observed in our transmitter is the effect of high hydrodynamic resistance, which causes the $w_\mathrm{p}$ to extend over time compared to the $T_\mathrm{g}$. This temporal extension is directly attributed to the properties of the mixing compartment, where increased resistance slows down the transmission of molecular signals. Despite this limitation, it is noteworthy that the relation between the applied $T_\mathrm{g}$ and the extended $w_\mathrm{p}$ remains linear, as shown in Fig. \ref{fig:tables} (b).

\begin{figure*}[ht]
    \centering
    \includegraphics[width=0.95\linewidth]{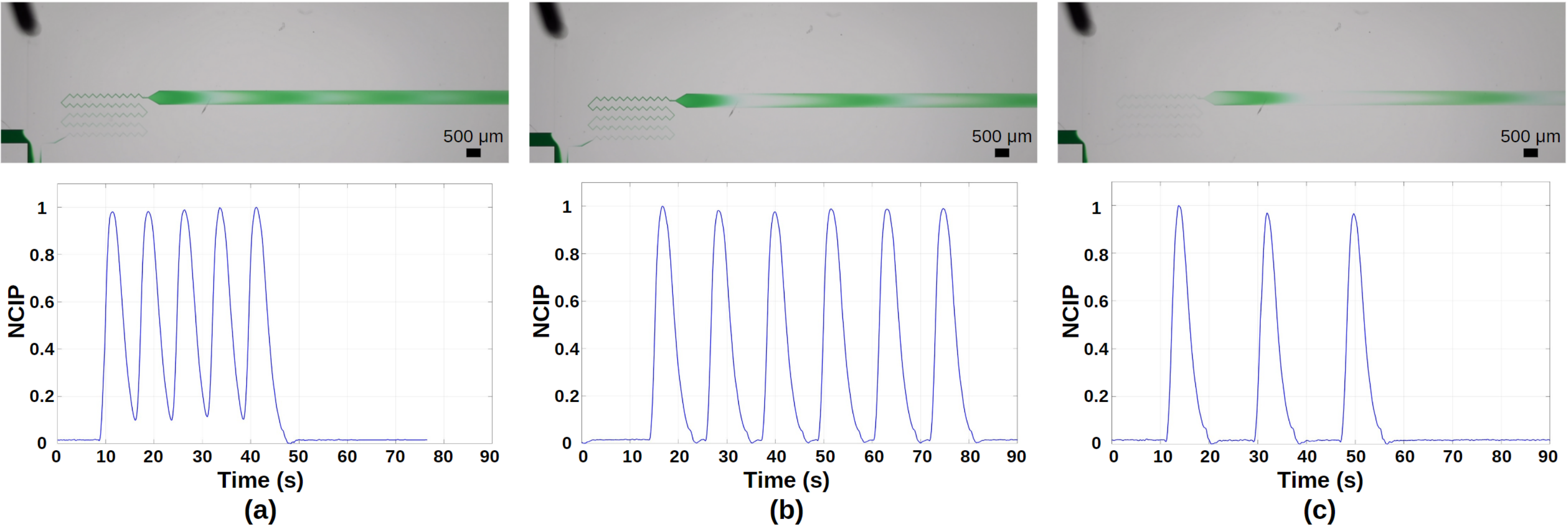} 
    \caption{An illustration of consecutive pulses with different input inter-pulse durations ($T\textsubscript{pi}$): (a) t = 7 s, (b) t = 11 s and (c) t = 17 s. NCIP stands for normalized concentration intensity profile. Processed with MATLAB.}
    \label{fig:IPD}
\end{figure*}

\begin{figure*}[h]
    \centering
    \includegraphics[width=0.95\linewidth]{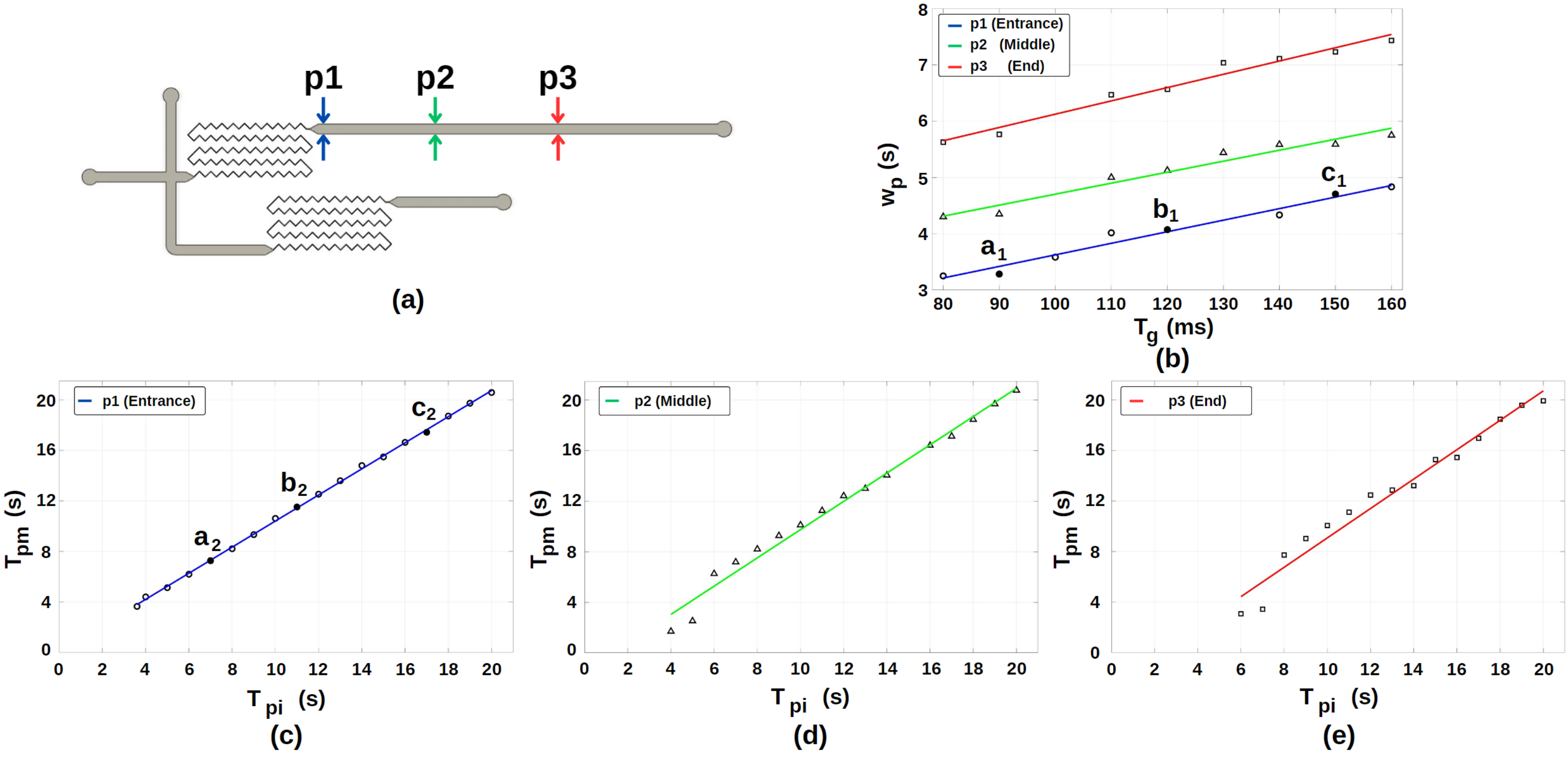} 
    \caption{Investigation of the impact of input parameters on transmitted pulses at three sampling points: p1, p2 and p3 (a). w\textsubscript{p}, for varying T\textsubscript{g} (b). T\textsubscript{pm}, for varying T\textsubscript{pi}; (c) at p1, (d) at p2 and (e) at p3. Corresponding images for data points a\textsubscript{1}, b\textsubscript{1} and c\textsubscript{1} (b); a\textsubscript{2}, b\textsubscript{2} and c\textsubscript{2} (c) are demonstrated in Fig. \ref{fig:PW} and Fig. \ref{fig:IPD} (Geometric-shape points represent the data points, while the colored lines denote the fit-line graphs).}
    \label{fig:tables}
\end{figure*}

In addition to varying $T_\mathrm{g}$, the investigation also explored the effects of different $T_\mathrm{pi}$ on the reliability performance of the microfluidic transmitter. Three different $T_\mathrm{pi}$ —$7$, $11$, $17$ seconds— are provided in Fig. \ref{fig:IPD}, to discuss the impact of $T_\mathrm{pi}$ on signal interference and the overall effectiveness of the system. A significant challenge leading to signal distortion and degradation in microfluidic communication systems stems from ISI due to the diffusion of molecules, dispersing the concentration along the propagation axis. Also, shorter $T_\mathrm{pm}$ can potentially increase the rate of information transfer, it also significantly elevates the risk of ISI, compromising the accuracy of the transmitted signals. This is evident in Fig. \ref{fig:IPD} (a), where interference effects become more pronounced with shorter $T_\mathrm{pi}$. However, our additional mixing structure addresses a critical aspect by ensuring homogeneously distributed concentration pulses across the cross-section of the channel. This precise control over concentration distribution helps mitigating the effects of ISI offering high spatiotemporal resolution at the propagation channel, and thus enables the development of new schemes.

Following the analysis of our transmitter’s consecutive pulse generation performance for three different input values of $T_\mathrm{g}$ and $T_\mathrm{pi}$, our investigation proceeds with the evaluations of $T_\mathrm{pm}$ and $w_\mathrm{p}$ at three sampling points, as illustrated in Fig.6 (a). These sampling points represent potential receiver locations that can be integrated into the microchannel. From the receiver design perspective, understanding the temporal dynamics at a fixed point is important, as it directly influences the receiver’s ability to accurately interpret the transmitted signals. Our experiments revealed a linear relationship between $T_\mathrm{g}$ and $w_\mathrm{p}$. Despite the dispersion of transmitted concentration pulses due to convection and diffusion, this linear relationship remained consistent, affirming the predictability of our design’s behavior. Furthermore, we explored the distinguishability of transmitted pulses by applying $T_\mathrm{pi}$ ranging from 3.6 s to 20 s. This examination showed not only a linear relation between $T_\mathrm{g}$ and $w_\mathrm{p}$ but also between $T_\mathrm{pi}$ and $T_\mathrm{pm}$. Moreover, the presence of a mixing compartment facilitated the generation of distributed flow along the cross-section of the channel, enabling precise control over signal characteristics. As a result,$T_\mathrm{pm}$ are approximately equivalent to $T_\mathrm{pi}$, as depicted in Fig. 6 (c). Despite the considerable distance (nearly 6 mm) between sampling points p1, p2 and p3, the majority of transmitted pulses tolerated the effects of ISI on distinguishability, thus maintained precision in $T_\mathrm{pm}$. However, certain pulses with smaller $T_\mathrm{pi}$ experienced dispersion, making them difficult to detect, as illustrated in Fig. 6 (d) and (e). This dispersion occurs because shorter pulse widths, associated with reduced gating times, lead to increased spreading of the pulses as they travel through the channel. More precisely, the reduced gating time allows fewer amount of molecules entering the propagation channel leading to a decrease in the concentration of molecules within the pulse and a greater susceptibility to dispersion effects. Nonetheless, the overall predictable behavior of our transmitter and the establishment of linear correlations between input parameters and measurements demonstrate its potential for integration into microfluidic testbed.

\section{Conclusion}
\label{sec:Conclusions}
This paper experimentally validates a previously proposed practical microfluidic MC transmitter based on the hydrodynamic gating technique, designed to overcome limitations in existing testbeds for chemical signal generation. With a cross-shaped architecture and zigzag mixing structure, our transmitter delivers precise, programmable, consistent, and reproducible molecular pulses with variable pulse widths and inter-pulse durations, facilitating advanced experimental research in MC. We demonstrate linear relationships between gating time and pulse width, as well as between applied and measured inter-pulse durations, confirming the reliability and predictability of our design. The additional mixing structure effectively mitigates ISI, resulting in concentration pulses that are distinguishable by a potential receiver. This platform provides a reliable testbed for MC system development, enabling further experimental research in MC and practical applications of IoBNT. Future work will focus on optimizing input parameters in this system to achieve tunable pulse amplitudes. Additionally, the system can be modified to transmit multiple molecular signal waveforms propagating simultaneously in the channel. Future work will include a quantitative performance analysis of this model by comparing experimental results with an updated version of the analytical model derived from our previous research.

\section{Acknowledgments}
This work was supported in part by The Scientific and Technological Research Council of Turkey (TUBITAK) under Grants \#120E301, \#123E516, and European Union’s Horizon 2020 Research and Innovation Programme through the Marie Skłodowska-Curie Individual Fellowship under Grant Agreement \#101028935. The authors also acknowledge the use of equipment procured through the AXA Research Fund (AXA Chair for Internet of Everything at Koç University - PI: Prof. Ozgur B. Akan), and the use of facilities of the n\textsuperscript{2}STAR-Koç University Nanofabrication and Nanocharacterization Center for Scientific and Technological Advanced Research.

\bibliographystyle{ACM-Reference-Format}
\bibliography{references}

\end{document}